# Material processing by laser-plasma-filament-guided high voltage discharges

Kristian Cvecek[1,2], Markus Döring[1,2], Alexander Romboy[1,2], Johannes Heberle[1,2] and Michael Schmidt[1,2]

[1] Institute of Photonic Technologies, Friedrich-Alexander-Universität Erlangen-Nürnberg, Erlangen, Germany
[2] SAOT – Erlangen Graduate School in Advanced Optical Technologies, Friedrich-Alexander-Universität Erlangen-Nürnberg, Erlangen, Germany

E-mail: kristian.cvecek@lpt.uni-erlangen.de

**Abstract**

We investigate ablation experiments performed by laser-plasma-filament guided electrical discharges at high-voltages of up to 145 kV. The guiding was accomplished via fs-laser-generated plasma filaments across a gap of 201 mm of air onto steel 1.3343 samples. This method combines remote material processing and enables the steering and deflection of high voltage discharges with the ease-of-use of remote laser processing technology. We observe an increase of the per-pulse-(and-discharge)-ablated volume by a factor of 1.67 over an ablation regime when the discharges are not present and up to a factor of 12.5 over the case when neither discharges nor plasma filaments, only a loosely focused laser beam, are present.

Keywords: ultra-short laser pulses, laser/plasma filament generation, laser-filament-guided high-voltage discharge, laser remote processing, remote electrical discharge processing

## 1. Introduction

As early as 1974 Leonard Ball ([1]) has recognized the potential of laser pulses to act as lightning rods by generating conductive plasma filaments (see for theory e.g. [2]) in air and the scientific community researching high power and ultra-short pulsed lasers has been fascinated ever since. Subsequently, many groups investigated this idea, be it to obtain non-straight plasma filaments and discharges [3], to optimize the onset of the plasma filament [4, 5] or the time duration for electric conductivity [6], culminating in several experimental studies for actual lightning capture in the field [7, 8].

While there are numerous other potential applications for plasma filaments in air, see [9], improving technologies for channeling lightning is only meaningful because lightning itself can be highly destructive due to the released electrostatic energy. This begs the question if something meaningful or useful (other than protection) can be done with a successfully grounded lightning, especially when its position for grounding is set beforehand by the laser beam.

While channeling natural lightning is highly impractical due to its extreme power and the unpredictable conditions within thunderstorms, the ability of directing high-voltage discharges offers nevertheless intriguing potential for remote materials processing. The underlying motivation is as follows: remote processing methods typically enable high processing speeds, convenient access to the workpiece and a straightforward automation. Among these, lasers are particularly well-suited, but their applicability becomes questionable if the required output power range is on the order of 100 kW or more. This limitation comes from the relatively low wall-plug efficiency of lasers compared to direct methods like electric resistance welding. Owing to the absence of intermediate energy conversion steps, direct electrical methods can reach up to 80% wall plug efficiency, while even state-of-the-art laser-diode based systems achieve only up to ~50% efficiency. Moreover, the investment costs for purely electrical devices with a given output power are often just a fraction when compared to a laser with comparable output power.

Our approach aims to bridge this technological gap by combining the advantages of laser remote processing with the efficiency of an electrical discharge. We use a rather "low powered" ultra-short pulsed (USP) laser to create a long, conducting filament in air and use it to guide a powerful electric discharge to process the workpiece at the position where the filament is pointing at.

While the actual implementation of the planned deflection scheme for the plasma filaments needs to be finalized yet (see Appendix A), we present here first results on how an electric discharge guided by a plasma-filament affects the ablation of a steel target. The structure of this paper is as follows: the experimental setup and conditions are described in Section 2. Section 3 presents the experimental results and is followed by Section 4, which discusses the results.

## 2. The Experiment

We use in our experiment a Ti:Sapphire based USP laser (Coherent Astrella-1K) capable of delivering laser pulses as short as 34.1 fs (FWHM) with up to 7 mJ of pulse energy at a repetition rate of 1 kHz and a center wavelength of 795 nm. The pulses can be pre-chirped to have a positive or negative chirp at the laser's output by adjusting the gratings inside the regenerative chirped-pulse amplifier (CPA), thus making pulse durations of up to 4.5 ps possible.

### 2.1 General setup

The laser is guided through several mirrors and a galvanometric scanner into an adjacent room that is equipped with a Faraday cage containing a Cockcroft-Walton high-voltage (HV) generator. The HV generator can deliver electrical discharges of up to 300 kV and can store up to 20 J of electrical energy, in case the electrical field strength at the electrode remains low enough during the charging cycle to prevent a spontaneous discharge at lower voltages. When entering the Faraday cage, the collimated laser beam has a $1/e^2$-diameter of 14.4 mm. Even though the laser pulses exceeds the pulse energy necessary for self-focusing, the size of the Faraday cage and the distance available for experiments were limited so the laser beam is pre-focused in order to generate suitably long filaments for the experiment at the right position.

To do so we used a lens combination with focal lengths of -2.5 m and +1 m to achieve an effective focal length of 1.667 m which gives a filamentation onset already after 1.45 m of propagation length, due the self-focusing of the laser beam. Due to the losses at the mirrors and the lenses, the pulse energy decreased to 4.4 mJ after passing the lenses. This led to a generation of visible plasma filaments with a length of up to 0.35 m (after the 1.45 m). However, as the filaments have different plasma densities along their length, their ends are barely visible by the unaided eye. To make sure that the electrode-workpiece distance in the presence of pulse energy fluctuations is not overextended, it is set to 0.21 m. Without Kerr-self-focusing the nominal $1/e^2$-laser spot diameter would be 130 µm at the sample, with a peak fluence of 49.8 J/cm² (per pulse). The pulses arriving at the sample after generating the filament are expected to retain at least 98% of their original pulse energy when irradiating the sample because the plasma filament generation does not take no more than approx. 4 µJ/cm of the pulse's energy [10].

Fig. 1 shows the schematic of the beam path in front of and inside the Faraday cage. The scanner motion, which was synchronized to the repetition rate of the laser, was chosen such that a series of three laser pulses (at 1 kHz) were deflected onto three mirrors – each pulse on its respective mirror and beam path. Together with the prisms in front of the lenses this design was used to emulate the generation of plasma filaments at different inclination angles between the electrode and workpiece (compare Appendix A). This particular feature of the setup was additionally used to determine the angle-dependent channeling probability. For the ablation experiments described here, only pulses from the center beam were used. The effective pulse repetition rate decreased due to the scanner setup down to ~ 23.9 Hz (from 1 kHz). As the HV generator needed to be manually triggered, single-pulsed laser shots were triggered manually when carrying out the discharge experiments. In this case, the duration between the manually triggered pulses/discharges was around 1 to 3 s.

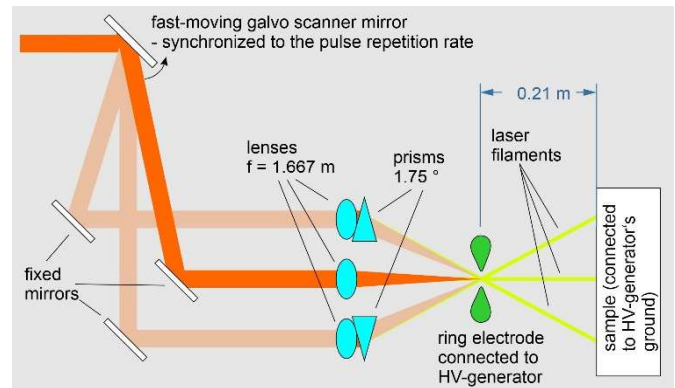

Figure 1: Schematic setup of the experiment.

Since the system was designed with full plasma-filament-deflection capability in mind, see Appendix A, we investigated also two different electrode shapes, type I and type II. The ablation experiments described here were conducted only with type II, because it shows a much higher capability for guiding electrical discharges through laser generated plasma filaments than type I. As this work primarily focuses on ablation experiments, Appendix B contains additional information on how the electrode shape affects the laser induced discharge guidability. Nevertheless, both electrode types were designed to be circularly symmetric and because of this, the plasma filaments needed to pass through a common pivot point inside the electrode.

## 2.2 Pre-compensation of GDD for optimal filaments

The onset position and length of the laser-generated filaments exhibits some day-to-day variation, which is unrelated to the laser parameters. We attribute these day-to-day variations mainly to variations in ambient humidity and air pressure (in the otherwise temperature-controlled lab), as water vapor exhibits absorption bands in the infrared range around 1.5 µm. This affects the dispersion and, consequently, the group delay dispersion (GDD) via the Kramers-Kronig relation [11], even at the laser's central wavelength of 795 nm, as the beam propagates for at least 7 m until it reaches the electrode.

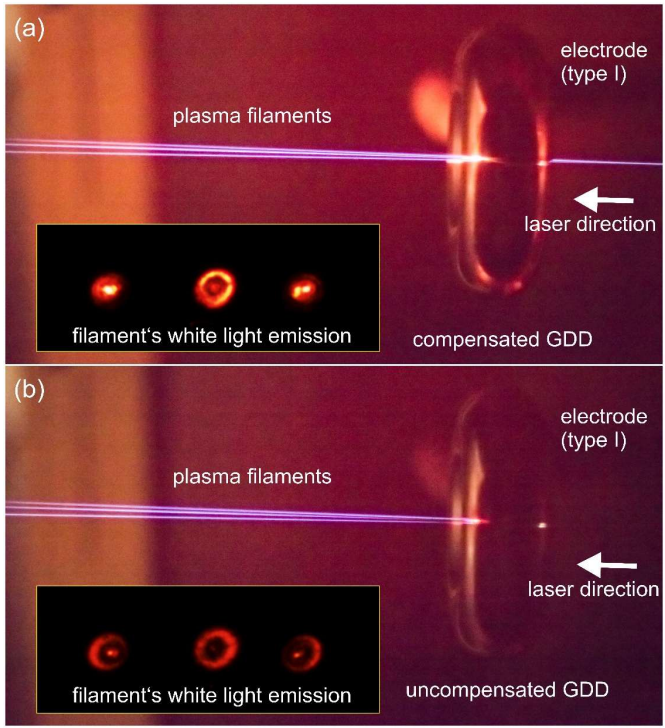

Figure 2: Effect of the pre-compensation of the GDD on the length and onset of the center and side plasma filaments. GDD pre-compensation is set so that all three filaments start before the electrode and extend beyond the left edge of the camera's field of view (a). Compared to (a) the GDD setting in (b) is unsuitable as it is not clear where exactly inside the electrode the filament begins. To make sure that the electrode is well "connected" to the sample via the filament, the filament must extend beyond the electrode to the right as in (a). .

Due to the variation of the GDD, the pulse duration changes, thus causing changes in the pulse's peak power, despite the constant pulse energy. In combination with the self-focusing in air and the optical components of the setup, this results in a change of the filament length and the filament's onset position. To keep the filament's length and onset position the same despite the observed variations, the gratings inside the regenerative CPA were adjusted accordingly.

Interestingly, the longest filament length and earliest onset were always achieved when the conical emission (CE) of blue-shifted light, known to originate from four-wave-mixing inside the filament [12–14], and the white light generated in the center of the beam had similar intensities at a camera observing the visible wavelength range, see Fig 2. When the GDD is set correctly, the filaments are longer and start earlier, Fig. 2 (a), which allows to check if the onset of the filaments is positioned sufficiently in front of the electrode's center.

## 2.3 Alignment and sample preparation

High-speed tool steel 1.3343 (HS6-5-2C), described in [15], was selected as the sample material due to its high wear resistance. Its elemental composition is almost identical to the AISI M2 steels which have a liquidus temperature of around 1670 K (solidus temperature around 1500 K) [16]. This choice of steel material enables the filament-guided remote discharge process to be evaluated under rather demanding conditions, thus more clearly revealing its performance differences compared to "conventional" USP-laser ablation. The grinded samples, 2.3 mm thick with a size of 25 mm x 75 mm, had an rms-surface roughness of 0.8 µm. They were mounted onto an aluminum plate with a size of 40 x 40 cm² and a thickness of 10 mm, which was connected electrically to the ground electrode of the HV generator. The overall electrical resistance was lower than 1 Ω. The aluminum plate had a slit made at the height of the plasma filaments (during the experiment the slit was completely covered by the samples), so that the laser beam and filaments could pass through, when needed for adjustment purposes such as pre-compensating the GDD, see Section 2.2. We ensured that there are no spontaneous unguided discharges at the planned distance and voltage settings for the experiment, see Appendix B, so that neither atmospheric changes nor any HV generator issues, e.g. ringing during voltage build-up or discharge, compromise the experiments.

## 2.4 The ablation experiment

The ablation experiments were done at a fixed pulse energy of 4.4 mJ at the voltages of 123, 140 and 145 kV. For each voltage setting, we carried out four irradiation series in which the samples were irradiated with 50, 100, 200 and 400 pulses with the respective guided discharges. The pulses were manually triggered, so the filaments are generated after the HV generator had time to fully recharge. The duration between the manually triggered pulses/discharges was around 1 to 3 s. The topology of the resulting craters was analyzed by a confocal laser scanning microscope (Olympus LEXT OLS4000) to determine the ablated volume and depth.

To separate the effect of the electrical discharges on the ablation from the effect of the laser, we also carried out ablation experiments, at the pulse repetition rate of 23.9 Hz provided by the scanner system, again with 3 experimental series at 50, 100, 200, and 400 pulses. These experiments were performed without the high voltage, but set such that in one case only the laser generated plasma filaments were present while in the second case the laser pulses were chirped so strongly that nearly no filaments were generated. At this setting the pulses had a pulse duration of 4.5 ps instead of the 34.1 fs. The significance of these settings was to distinguish between the conditions where the interaction occurs between the sample and a laser generated plasma filament on one hand and a "conventional" laser pulse without causing filamentation on the other hand. The ablated volumes, depths and crater sizes were analyzed. For further details on the determination of the crater size see Appendix C.

We will refer to the ablation modes as follows:
(i) *discharge mode,* when the laser generates a plasma filament and guides and triggers an electrical discharge
(ii) *filament-only mode*, when the laser generates a filament but no discharge occurs (voltage set to zero)
(iii) *no-filament/no-discharge mode,* when neither a filament nor a discharge is present.

## 3. Results

An exemplary overview (white light microscope) of the crater morphology for all three cases (*discharge, filament-only* and *no-filament/no-discharge modes*) for an exposition to 400 pulses is shown in Fig. 3. The overall size of the crater in the *discharge-mode* is larger than the craters for the other modes. There is also significant change in the size of the crater: the *discharge mode* provides a large elliptical interaction area on the sample (Fig. 3 (a1)), while the *filament-only-mode* exhibits a more localized interaction, which is, however, not as circular as the *no-filament/no-discharge mode*, see Fig. 3 (c1).

In addition, the crater morphology for ablation with *filament-only* and *no-filament/no-discharge modes* shows at the bottom structures similar to cone-like protrusions (CLPs) that are known to form during the irradiation of metals with fs-laser pulses [17, 18]. The absence of CLPs in Fig. 3 (a1, a2) shows that the electrical discharge either interrupts or masks the process necessary for their formation. Moreover, a second striking difference of the *discharge mode* is the apparent lack of significant oxidation in the middle of the crater, as the crater ground exhibits a metallic sheen of the crater surface, whereas its edges exhibit annealing colors ranging from yellow to blue. In contrast, the other modes exhibit craters with a significant darkening.

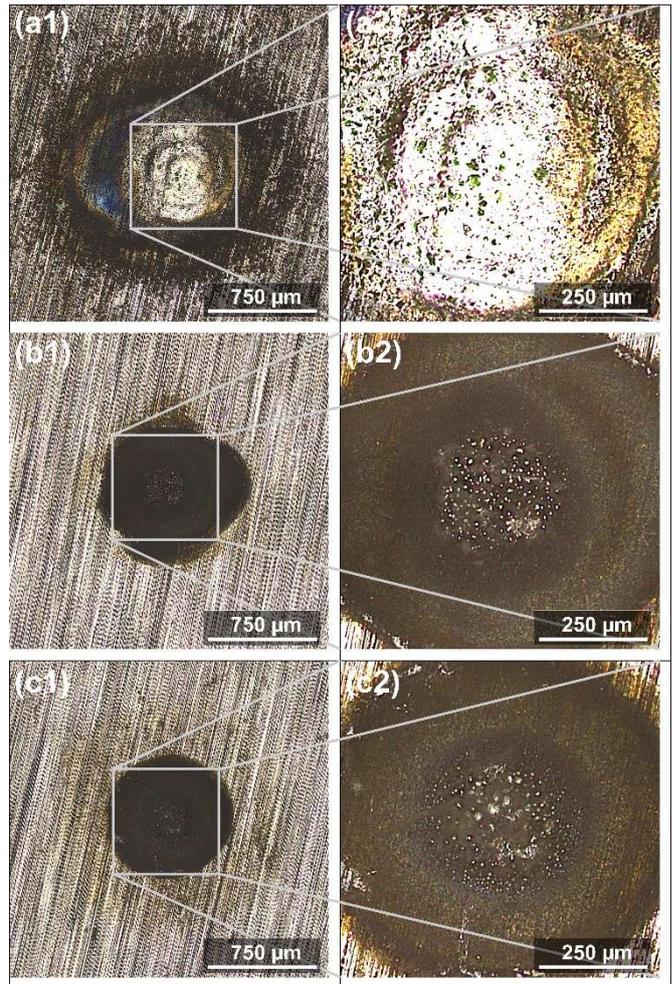

Figure 3: Crater morphologies for *discharge-mode* processing at 145 kV (a1), the *filament-only mode* (b1) and *no-filament/no-discharge mode* (c1). The right side shows zoom-ins of the corresponding inner crater regions with increased brightness and contrast (a2, b2, c2).

The measured crater topologies are exemplarily shown in Fig. 4 for all investigated modes and each applied pulse number. As can be seen, for the *discharge mode* the depth of the crater evolves monotonically deeper, without any "sharp" features, while for the other ablation modes, CLP-like structures start to appear. They become already at 100 applied pulses so pronounced that their height actually exceeds the surface level of the sample.

In order to quantize the properties the ablated craters we measured the effective ablated area (calculated from the crater morphology below surface level, see Appendix C), the effective ablated volume and the maximum achieved ablation

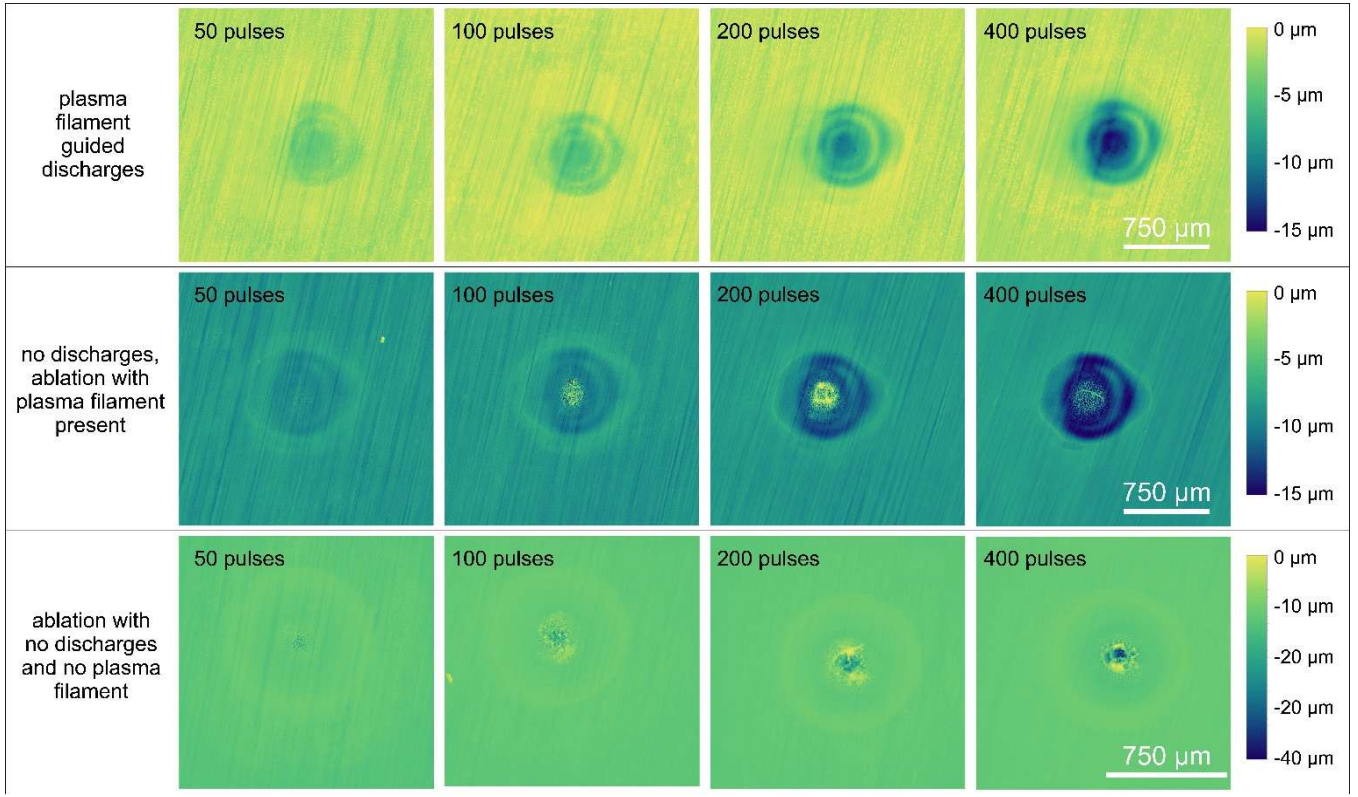

Figure 4: Exemplary crater depth morphologies for the guided discharges at 145 kV (top row), no discharges but laser generated filament present (middle row) and no discharges and no filament (bottom row). The numbers in the left upper corner show the applied number of pulses (and discharges when applicable). Please note that the lateral scale bar and the height scale differ for the bottom row from the upper rows.

depth, which is not necessarily at the exact geometric of crater, all shown together in Fig. 5

Using the appearance of the craters under a white light microscope as a measure for crater size, compare Fig. 3, we see that the *filament-only* and the *no-filament/no-discharge modes* have more characteristics in common than with the *discharge mode*. Both crater grounds exhibit CLP-like structures, which seem to cover comparable dimensions, while the area, where the debris has significantly darkened the surface is also comparable in size.

In contrast, when defining the crater area as the region within a rim that is defined by the height of the crater decreasing below the sample's surface height, quantitatively far more similar results are obtained for the *discharge* and *filament-only modes* with effective ablation areas between 0.5 and 0.8 mm². This is shown in Fig. 5 (a). When comparing the area of the *filament-only mode* to those of the three *discharge modes* at the respective pulse numbers (averaged across different voltages) we find it to be on average about 83 % of the size of the *discharge modes*. We believe averaging the *discharge modes* data is possible here because for pulse numbers of 200 and lower, their standard deviations overlap. In addition, the individual voltage settings do not show a clear trend. Instead, the 140 kV voltage yields the same area for 200 pulses as the voltage at 123 kV, while being different for 400 pulses. However, the discharge energy of 140 kV is much closer to 145 kV than to 123 kV, and should result in a larger area, if equating higher discharge energy with larger interaction areas would be valid. Therefore, we believe that the interpretation of the ascending ordering of the *discharge modes* at 400 pulses with increasingly higher voltages should be considered a random outcome rather than an interpretable result despite the non-overlapping standard deviations. For clearer results, further measurements would need to be conducted at much higher pulse numbers or at a larger range of discharge voltages. Nevertheless, both *modes* are in stark contrast to the *no-filament/no-discharge mode* as this mode exhibits an area approximately 50 times smaller than for the other two modes.

Fig. 5 (b) shows the ablated volume for all investigated modes. The ablated volume is significantly larger for the *filament-only* case of the plasma filament when compared to the *no-filament/no-discharge mode*. This is surprising, because the generation of the ionized plasma actually decreases (even if negligibly) the available pulse energy that can fuel the ablation on the sample.

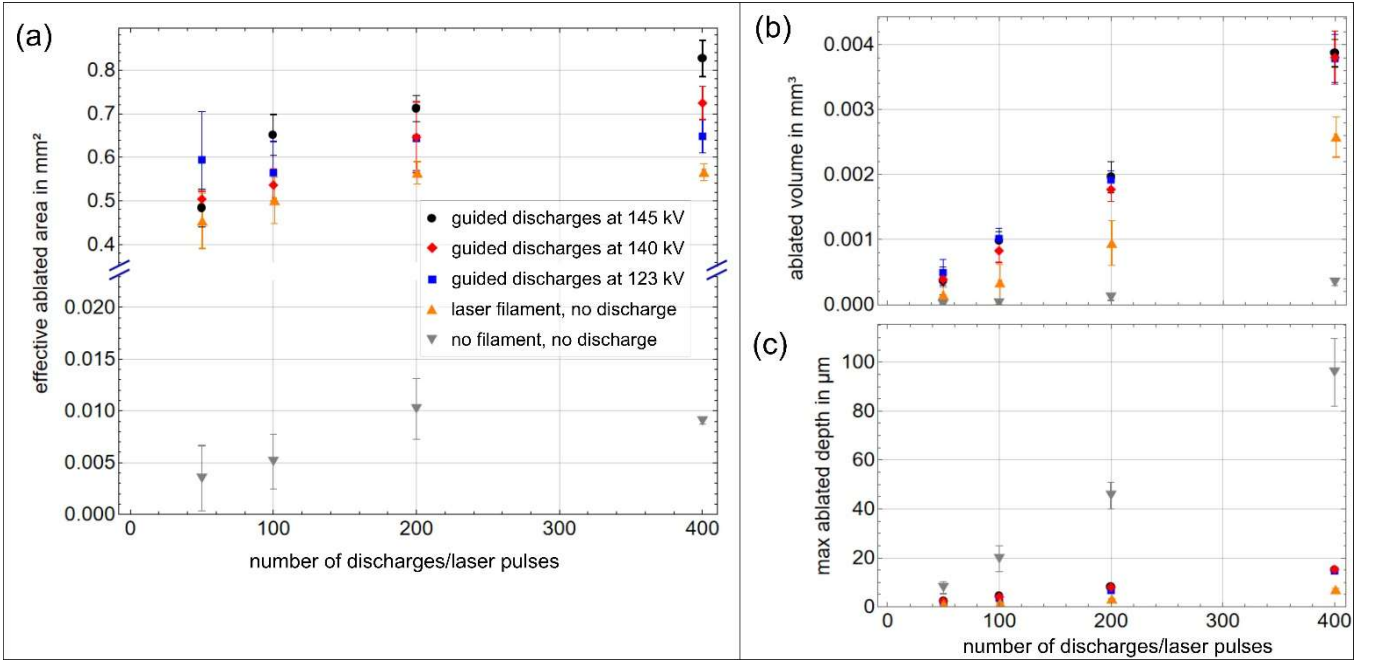

Figure 5: Average ablated area determined for crater heights below the sample's surface level for all investigated modes – (a) (note the break of the y-axis). Average ablated volume for all investigated modes – (b). Maximum achieved ablation depth for all investigated modes – (c).

While the *discharge modes* provide the largest ablated volume, the discharge voltage does not appear to have a significant effect on the ablated volume. With the discharge voltages lying in a range between 123 to 145 kV, the highest voltage should provide nearly 40% more discharge energy than the lowest voltage according to the energy of a charged capacitor. Nevertheless, the relative standard deviation lies in a range between ±5 % to ±9 % for all three different discharge voltages. Consequently, there is a discrepancy between the observed and expected ablation volumes, when assuming that the discharge voltage contributes directly to an increase of the ablation volume.

Fig. 5 (c) shows the maximum achieved depth during the process. Note, that the maximum depth reported here does not necessarily correspond to the geometric center of the crater (see Figs. 3 and 4). Clearly, the depth of the *no-filament/no-discharge mode* exceeds the other modes by a large margin. However, similar to Fig. 5 (b) showing the ablated volume, the *discharge mode* provides greater depths than the *filament-only mode*. When comparing Figs. 3 and 4, there is also significant remelting present, which is known to occur during drilling of metals with ultrashort laser pulses [19]. For a better overview of the results here, the main findings of this section are summarized in Table 1.

| | electrical discharge (DM) | filament only (FOD) | no-filament/no-discharge (NFND) |
|---|---|---|---|
| white-light microscope appearance of craters (Fig. 3) | • largest size<br>• elliptical, irregular rim<br>• metallic sheen with tempering colors | • smaller size than DM<br>• somewhat irregular rim<br>• cone-like protrusions<br>• dark crater ground | • size similar to FOD<br>• very regular (round) rim<br>• cone-like protrusions<br>• dark crater ground |
| depth map (Fig. 4) | • crater depth ~ pulse no.<br>• no features above sample's surface | • crater depth ~ pulse no.<br>• features above sample's surface ~ pulse no. | • crater depth ~ pulse no.<br>• features above sample's surface ~ pulse no. |
| effective ablation area (Fig. 5 (a)) | • largest but similar to FOD (0.83 mm²)<br>• inconclusive discharge voltage dependency | • 83 % of DM<br>- | • 50x smaller than DM<br>- |
| ablated volume (Fig. 5 (b)) | • largest (0.0038 mm³)<br>• inconclusive discharge voltage dependency on pulse no. | • 60 % of DM<br>- | • 8 % of DM<br>- |
| max ablated depth (Fig. 5 (c)) | • 16% of NFND<br>• inconclusive discharge voltage dependency on pulse no. | • 5% of NFND<br>- | • largest (95 µm)<br>- |

Table 1: Summary of the main results. The different investigated modes, i.e. discharge, filament-only and no-filament/no-discharge modes *are abbreviated as DM, FOD, and NFND in the table.*

## 4. Discussion

It is clear that the observed differences in the results must be explainable by the different mechanisms each mode brings to bear on the sample. Therefore, before discussing the possible causes for the observed results it is instructive to recall these mechanisms and draw a timeline at which they apply. This is done, based on Tzortzakis et al. (2001) [20] and Agnihotri et al. (2017) [21], for air/laser-filament and air/electric-discharge interactions, respectively, in Fig. 6. In addition, the typical time ranges when phase transitions of the laser irradiated material occur are shown alongside.

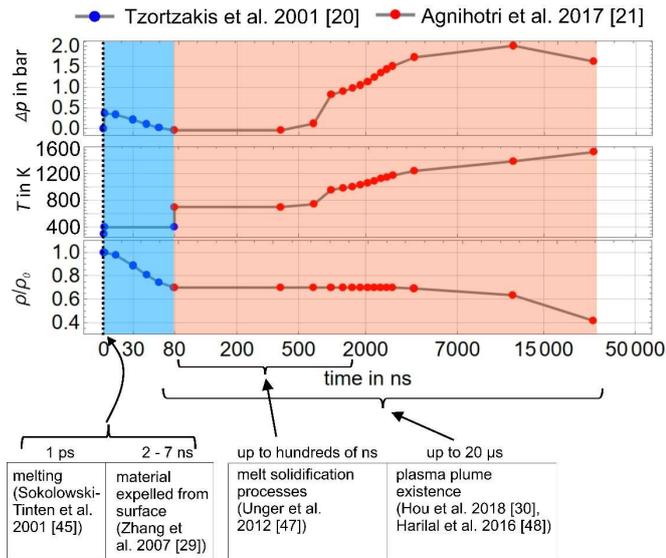

Figure 6: Estimation of the thermodynamic evolution of air during a laser triggered discharge all investigated modes.

Note, that Tzortzakis et al. [20] describe only the effect of a fs-laser pulse on the density of a plasma filament and not the pressure. We estimate for the discussion here the pressure by using the density by values and temperature rise (within the first 1 ns) given by [20] using the ideal gas law. In addition, the values for temperature, density and pressure evolution given by Agnihotri [21] are just for an electric discharge triggered by free-floating charges in air at normal conditions. Therefore, we assume in Fig. 6 that the effect of the electric discharge can be added to the effects of the laser pulse generating the filament at the delay of 80 ns, when the air density within the filament drops far enough to sufficiently reduce the air's breakdown voltage to trigger the electric discharge due to the Paschen effect [20].

Although the parameters in all publications Fig. 6 is referring to differ from each other and from our experimental settings, we nevertheless believe that Fig. 6 can help explain our results. It shows effectively that the melt and plasma plume formation happen on a picosecond timescale while the electric discharge occurs far later, when the plasma plume has already formed and portions of the melt remain at the irradiated area. Therefore, the electric discharge might affect the re-solidification process but not the actual ultra-short-pulse-triggered ablation.

### 4.1 Ablation volume increase of the filament-only mode

We start our discussion with the, in our opinion, most interesting result from Section 3. With this, we mean the fact that the ablated volume in the *filament-only mode* is 7.5 times larger than the ablation using the *no-filament/no-discharge mode*, where all the pulse energy is actually available for ablation, while using the *filament-only mode* a portion of the pulse energy is lost due to the generation of the plasma filament. Nevertheless, as pointed out by Rosenthal et al. [10], this amounts only to 2 % losses in our experiment at maximum, (compare Section 2.1).

While it is known that the (per-pulse-specific) ablation efficiency decreases for too high pulse energies, see e.g. [22, 23], the decrease is unlikely to exceed a factor of 3, even if the applied laser pulse fluence of nearly 50 J/cm² is far larger than the optimum fluence, see Section 2.1. In contrast, we observe here a factor of 7.5, which is much higher than the factor of 3 by this effect. This is further compounded by the fact that through self-focusing and plasma action of the plasma filament, the filament has a diameter of approx. 100 μm [2], while the *no-filament/no-discharge mode* has nominally a focal spot of 130 μm at the sample (see Section 2.1). The beam size (by a factor of 1.69) of the *no-filament/no-discharge mode* suggests that its fluence should be by a factor of 1.69 smaller on the sample, and should thus be closer to the optimum fluence. However, this is clearly not the case as it has a significantly lower ablation efficiency than the *filament-only mode*.

In both cases, the electrical discharge does not occur, and therefore the only difference is the presence or absence of the laser generated plasma filament. Consequently, the interaction between the plasma filament and the irradiated region must be responsible for the efficient removal of material. One important effect, which may explain this observation is the so-called background energy reservoir [24]. The energy reservoir is basically formed by portions of the pulse energy being refracted into the adjacent (unionized) air via the plasma core, and being self-focused back into the filament in a dynamic balance [13, 25]. As this reservoir can contain up to 90% of the pulse's energy [26] and have at least a size of 300 μm in diameter [27], an ablation of the sample is highly probable, and would explain the large interaction area. Moreover, the acting fluence would be significantly reduced, consequently increasing the ablation efficiency. Indeed, calculating the energy specific ablation volume we obtain 1.5 μm³/μJ, compare Section 4.3, which is similar to values reported by Winter et al. [22] for fluence values no too far away from the fluence optimum.

In addition, as illustrated in Fig. 6, the generation of the plasma filament leads to a temperature rise of about 100 K above room temperature, with subsequent air density reduction down to 70 % [20]. In an earlier work, Tzortzakis et al. [28] show that during the plasma filament generation the air is only partially ionized with electron densities up to $2\times10^{17}$ electrons/cm³, while air has at normal conditions a molecular number density of approx. $2.7\times10^{19}$ molecules/cm³. Moreover, the recombination of the majority of the ionized particles should be finished within 1 ns, leading to the already mentioned temperature rise about 100 K [28]. Because the material lift-off from the sample surface occurs at picosecond time scales (compare Fig. 6, [29]), the ablated material might readily react with the present reactive plasma.

The slight increase in air pressure due to the filament formation might not be effective in a better removal of the ablated material. This is because the ablated material's shock front has a much higher momentum than the air's pressure increase can cause, especially when at time instances at tens of nanoseconds and later the penetration of the ablated material deeper into adjacent space is made easier due to the lowered air density. This is corroborated by [30], who show that the ablation plume propagates much further into the generated filament than when no filament is present. This way, the amount of material available to recondense onto the sample might be reduced in comparison to the *no-filament/no-discharge mode*.

Moreover, the effective crater area and the crater shape between both modes evidence the significantly better material removal in the presence of the plasma filament. The *filament-only mode* shows CLPs in Figs. 3 and 4, as does also the *no-filament/no-discharge mode*, but the *filament-only mode* lacks a rim of recast melt at the edges of the crater, which is, conversely, present for the *no-filament/no-discharge mode*. The presence of such a rim suggests that there is significant motion of the melt, which solidifies before it can leave the sample [31]. However, as this rim does not exist during the *filament-only mode* there must be processes involved, such as described above, which improve significantly the material ablation.

In addition to the action of the plasma filament on the re-condensation, there exist several other possible effects that can increase the ablation when a plasma filament is present. As shown in Section 3, the ablation crater for the *filament-only mode* shows irregularities in its shape in comparison to the *no-filament/no-discharge mode* (see Fig. 3). These might stem from a nonlinear beam pointing jitter induced by filamentation [32, 33]. In addition to the beam pointing jitter there might be also light impinging on the sample in an off-axis fashion, which is known to originate from the conical emission (CE) discussed in Section 2.2. One clue to this are the ring-like structures visible in Figs. 3 and 4. While the intensity of the CE is far weaker than the light propagating on axis, the CE's spectral components will cover a very large portion of the white light, potentially reaching into the UV range. Due to the coherent nature of their generation these components constitute nevertheless ultra-short pulsed radiation. Because of this and the shorter wavelength, they might nevertheless significantly contribute to some of the observed off-axis ablation despite their lower intensity. This is even more likely, because during the *no-filament/no-discharge mode* there is no plasma filament present and hence also no CE [14].

Further work is necessary to determine which of the above effects contributes predominantly to the significantly larger ablation volume: the background energy reservoir, the decreased air density from the plasma filament, nonlinear beam pointing, or CE. Nevertheless, even a slight broadening of the crater, which makes it shallower, will increase the ablation efficiency. This is because at highly oblique angles the ablation threshold rises as shown by [34].

### 4.2 Mechanisms acting during the no-filament/no-discharge mode

Now that we have discussed several effects that can cause the *filament-only mode* to exhibit so much higher ablation efficiency compared to the *no-filament/no-discharge mode*, it is interesting to analyze why the *no-filament/no-discharge mode* is so much smaller in ablation volume and effective ablation area, but yields the largest achieved maximum depth.

We can expect that the impinging pulses energy will initially lead to material ablation by phase explosion, where the irradiated material is overheated above "[…] the limit of thermodynamic stability of the liquid layer […]" [23] for both modes. Even if the fluences are below the threshold for phase explosion, spallation will occur, which is characterized by inertial stress confinement caused by the incapability of the sample's material to expand as the electrons that have absorbed the laser pulse's energy thermalize with the lattice. Consequently, the region is put under high compressive stresses, which relax through the "[…] generation and growth of sub-surface voids which, in turn, may cause separation and ejection of liquid surface layers or droplets from the bulk." [23]. Effectively, in both cases material from the irradiated area is expelled, which is either liquid or cools with time down to become a liquid. This can cause even for single pulse ablation an increased rim of the ablation crater [35], as we observe e.g. in Fig. 4 for the *no-filament/no-discharge mode*.

However, if the ablation crater grows deep enough, the expelled material can adhere to the crater walls while starting to clog and shadow the entrance pupil to the crater [36]. Thereby it will reduce the available average intensity in the deep part of the crater. Since for the *no-filament/no-discharge mode* there are no debris removal processes going on, while at the same time there is no background reservoir, no CE and no nonlinear beam pointing jitter, the ablation of the area adjacent to the crater will be ineffective. Only the region inside, at which the pulse has a sufficiently high intensity will continue to be ablated.

One other possible reason for the observed crater morphology of the *filament/no-discharge mode* might be caused due to the pre-chirp of the pulses used to suppress plasma filament formation. In order to suppress the filament, the pulse duration has been increased up to 4.5 ps, which is the maximum the chirped-pulse amplifier of our laser can provide.

For pulse durations below 1 ps, the ablation fluence threshold increases with the pulse duration raised to the power of 0.05 [37]. For durations longer than 1 ps, it scales with the square root of the pulse duration [38],[39]. Overall, this corresponds to an approximate 2.5-fold increase in ablation fluence between 34 fs and 4.5 ps pulses, giving about 20 % from 34 fs to 1 ps, and a factor of 2.1 from 1 ps to 4.5 ps, resulting in the 4.5 ps pulses exhibiting roughly 2.5 times higher ablation fluence than the 34.1 fs pulses used in the other ablation modes. In addition to the increased threshold fluence, the longer pulse duration would provide also a lower energy specific efficiency (ESAV) [22]. Therefore, the main differences of the *filament/no-discharge mode* to the other modes are very likely the reduced ESAV in combination with the increased ablation threshold, the lack of the background reservoir, and no additional debris-removal processes.

*4.3 Mechanisms acting during the* discharge mode

It is particularly striking that, despite the discharge voltage increasing from 123 kV to 145 kV, equivalent to nearly a 40 % increase in discharge energy, the ablated volume, effective area, and maximum depth remain essentially unchanged within the experimental uncertainty (see Fig. 5 and Table 1). Compared to the *filament-only mode*, the electrical discharge can add at most 3.3 to 4.7 J to the ablation process, as these are the energies of a charged capacitor calculated for this case. However, not all of the energy of the capacitor will reach the sample through a discharge, as the electrical discharge will fuel (an additional) ionization of the plasma channel, while another portion of the discharge energy will dissipate as heat, sound and radiation from the plasma channel.

The effective ablated area of the *discharge mode* is approx. 20 % larger than the *filament-only mode*. This means that the crater's surface slope is comparable in terms of effective intensity acting on the surface, see [34]. This allows us to obtain a rough estimate on the necessary energy for ablation in the *discharge* and *filament-only modes*, which in turn gives us an estimate on the energy efficiency of the electric discharge.

To obtain this, we approximate the sublimation enthalpy of our 1.3343 steel samples in Appendix D to be ca. 63 J/mm³. With this, the energy cost is at least 0.41 mJ/pulse (*filament-only mode*), when calculated for the ablated volume of 0.0026 mm³ observed at 400 pulses. As the laser pulses have an impinging pulse energy of 4.4 mJ, including plasma filament losses of 2%, see Section 2.1, only 9.5 % of the pulse energy drive effectively the ablation.

Nevertheless, calculating the energy specific ablation volume (ESAV) with these values we obtain 1.5 µm³/µJ. This is highly comparable to ESAV values reported by Winter et al. [22], as they used AISI 304 steel, which does not contain tungsten and 525 fs pulses (as opposed to our 34.1 fs pulses). The low efficiency in our case is readily explained by the reflectivity of the material, e.g. 50 % at high fluences for steel [22], and the process dynamics of heat conduction, radiative losses, shockwave formation and residual melting, compare [23]. To be more precise, the laser peak pulse fluence used here (~50 J/cm²), is by a factor of at least 50 larger, while the typical ablation fluences for steels are far below 1 J/cm² for fs-pulses [40],[22]. Consequently, our experiment is far above the optimal ESAV, and runs thus at a significantly decreased ablation efficiency, unless the effective fluence is reduced by the formation of the background energy reservoir due to the plasma filament, as discussed in Section 4.1.

Now, when we conduct the energy cost consideration for the ablation volume of the *discharge mode* (0.0038 mm³), we obtain 0.6 mJ/pulse. The difference energy of 0.19 mJ/pulse to the *filament-only mode* must be supplied by the additional electric discharge. When we assume a drastic (=linear) simplification that the electric discharge drives similar processes as an impinging pulse at the sample (heat dissipation, radiative losses, shockwave formation, etc.), we could expect a linear dependence of the effective discharge energy and the used discharge energy. Assuming that the proportionalities are similar for the electric discharge, the 0.19 mJ/discharge would necessitate at least 2 mJ of discharge energy. When we take into account that electrons are absorbed by a metallic workpiece (connected to ground) with nearly unity than this energy would be reduced to 1 mJ. However, the discharge takes place at µs time scales [41], [21]. At such time scales, the parasitic processes such as heat conduction are much more dominant and give rise (in the case of pulse laser ablation) to ablation thresholds that lie several orders of magnitude above those for pulse durations below 1 ps. In fact, Endo et al. report for copper a fluence increase of factor 1000 for pulse durations between 1 ps and 1 µs [38]. Consequently, the impinging electrical discharge energy on the sample can be speculated to be around 1 J.

This estimation is necessary, if we want to understand the observation of nearly-similar ablation volumes despite the increase of 40 % in the discharge energy due to the applied voltages. As already reasoned above, the discharge energies lie for these voltages between 3.3 to 4.7 J. If we assume that a 3.3 J discharge provides the estimated 0.19 mJ/discharge at the sample, linear approximation would yield a value of 0.27 mJ/discharge for a discharge energy of 4.7 J. Since the discharge occurs after most laser-pulse-induced processes have abated or even stopped, we can add both energies. This gives e.g. for the discharge of 3.3 J an effective ablation energy of 0.6 mJ/(pulse&discharge) and for the 4.7 J discharge an effective 0.68 mJ/(pulse&discharge). The average of both values is 0.64 mJ/(pulse&discharge). The relative difference of both values to the average one is thus ±6.25 %, which might explain the observed standard deviations of the measurements for each individual voltage to be between ±5 % to ±9 %.

Of course, these considerations use several severe simplifications. However, without further experimental investigations, the uncertainties accompanying these simplifications cannot be resolved: While there exist several works discussing the different energy contributions during an electric discharge, the results are not readily usable for the understanding of our case. On the one hand, there are publications that investigate high voltages up to 1 MV at large distances (> 1 m), e.g. [41, 42], but they aim for the understanding of atmospheric discharges and therefore do not elaborate on the energies that can be transferred to an anode as in our case. On the other hand, there are works studying the transferred and deposited energy of a spark discharge, e.g. [43, 44] but they are limited to few kV, gaps around 1 mm and typically at high pressures. Last but not least, the actual discharge duration might be a crucial factor governing the ablation efficiency.

Besides the observed low sensitivity of the ablated volume on the individual discharge voltage, there are other noteworthy, unique features of the *discharge mode* craters, such as the polished look and the tempering colors (Fig. 3 (a)). Clues for an explanation of this may be found in Fig. 6. It is clear that the onset of the electrical discharge happens around 80 ns [20], when most of the material is already being expelled from the interaction area [29], since the melting of the material starts as early as 1 ps [45]. While König et al. [46] observe at 100 to 1000 ns a second transmission decrease, which they attribute to a secondary ablation from the sample, the motion of the free electrical charges starts to significantly heat the air at the end of this time scale (1000 ns) according to Agnihotri et al. [21], corresponding to 1080 ns in Fig. 6. Obviously, at this time instant, the interaction of the material and the laser pulse has already run its course and the discharge will affect only the ensuing processes on the sample.

When now the electrical charges start to heat and ionize the air, the pressure of the discharge channel starts to rise significantly. We believe that at this time, due to the high pulse energy there is still at least a thin layer of melt remaining in a liquid state on the crater's surface, as the complete re-solidification can take up to nearly 2 µs [47]. Without the electrical discharge the melt would actually form cone-like protrusions (CLPs) as in Figs. 3 (b) and (c), but the blast from the expanding gas of the discharge channel disrupts this self-organizing process to a large degree. So far in fact, that, completely opposed to the *filament-only mode*, the surface level of the crater surface lies always below the level height of the pristine surface region around the crater, compare Fig. 4.

Then, at 5 to 15 µs the temperature of discharge channel reaches a maximum, while its (particle) density starts to decrease even further than induced by the laser filament. At this time moment, there might be still ablation debris present [30, 48], and the low density of the discharge channel allows it to distance itself further from the sample and react more thoroughly with the hot molecules and ions of the ionized discharge channel. Because of this, there is virtually no debris inside the crater center, while only the far edges of the crater, where the discharge channel temperature is low and the air density high some of the debris accumulates.

Furthermore, the exposure of the sample surface at the edges of the ablation crater, where there is no melt or where the melt has already resolidified (basically the area around the CLPs), is exposed to very high temperatures for at least tens of microseconds, compare Fig. 6. The repeated exposure to such high temperatures, even if just for microseconds, will produce the observed tempering colors for the *discharge mode* in Fig. 3 (a). Unfortunately, we do not have the equipment to fully investigate the thickness and (phase) composition of the oxidation layer to explain the observed coloring of the interaction zones.

Nevertheless, because of the large time delay between the irradiation of the sample and the formation of the electrical discharge, all processes specific to the *filament-only mode* that are discussed in Section 4.1 should be present during the *discharge mode* too. These processes are the larger irradiation on the sample due to the background energy reservoir, better removal of debris caused just by the presence of the plasma filament (compared to the *no-filament/no-discharge mode*), nonlinear beam pointing jitter and a larger ablation size due to conical emission (CE).

However, what is striking for the interaction zones of the *filament-only mode* in Fig. 3 (b) and the *no-filament/no-discharge mode* Fig. 3 (c) are the similar dark colorations and surface topographies around the core region of the interaction zone despite the fact that for the *filament-only mode* a significant amount of material is ablated from the whole of the dark region while for the *no-filament/no-discharge mode* basically no material was ablated outside the direct interaction zone at the beam's axis.

Besides possibly resulting from recondensed material, this could be also explained by the properties of the oxide layers forming during solidification. For steel it is reported that the cooling conditions can have a strong influence on the properties of the resulting oxide layer [49, 50]. The different cooling conditions of *discharge mode* and the modes without the discharge could explain the difference in coloring. On one hand, the *discharge mode* in Fig. 3 (a) would form an oxide layer showing the typical tempering colors due to a (suspected) lower cooling rate because of its higher energy input. On the other hand, the dark coloration of the *filament-only* and *no-filament/no-discharge modes* may be caused by higher cooling rates resulting in the formation of metastable oxide phases with dark coloration and a high defect density of the oxide lattice. This, however, would also need to be studied closer for a definitive conclusion.

## 5. Summary and outlook

We have conducted ablation experiments on steel using a combination of ultra-short laser pulses and electrical discharges, where the ultra-short laser pulses generate a (in principle steerable) plasma filament, which triggers a high voltage discharge from a Cockcroft-Walton generator of up to 145 kV. For comparison, we conducted the ablation experiment using laser-filament triggered discharges (*discharge mode*), using laser pulses without an electrical discharge while generating a plasma filament (*filament-only mode*) and using neither an electrical discharge nor a plasma filament (*no-filament/no-discharge mode*). We carried out the experiment for different numbers of pulses and at different discharge voltages. The ablation volume, the maximum achieved depth and the effective ablated area were measured.

The results show that the ablated volume for the *discharge mode* exceeds the *filament-only mode* by about 67 % and is by a factor of 12.5 times larger than the *no-filament/no-discharge mode*. For the effective ablated region, defined as the rim of the crater below the (not irradiated) sample's surface height, the *discharge mode* exceeds the *filament-only mode* by about 20 % and is by a factor of 50 times larger than the *no-filament/no-discharge mode*. Only for the maximum ablated depth (not necessarily found in the absolute center of the ablation crater), the *filament/no-discharge mode* was higher by a factor of 6.25 compared to the *discharge mode* and by a factor of 20 compared to the *filament-only mode*. While the effect of the electrical discharge through the plasma filament clearly increases the ablation volume compared to the other modes, the discharge-material interaction has shown itself to be relatively insensitive to the investigated individual voltage settings of the discharges.

We discuss several potential interaction mechanisms that might contribute to the observed results in Section 4. Note, that our work here has solely focused on the ablation experiments. Therefore, the discussion of the interaction mechanisms is based only on reference literature that does not necessarily represent our experimental situation and conditions perfectly. Therefore, additional work is needed to evaluate and verify our hypotheses.

Nevertheless, the main mechanism we believe to increase the ablation volume for the *filament-only mode* so significantly above that of the *filament/no-discharge mode* is the presence of the plasma filament, which creates a background energy reservoir of the impinging laser pulse, reducing effectively the impinging fluence on the sample, while increasing the interaction area. There may be additional processes that lead to an increase of the ablation efficiency, such as the reduced air density due to the formation of the plasma filament or nonlinear beam pointing jitter. While the contribution of the guided electrical discharge is significant, it is by far not as strong as the effect of the plasma filament. This is because only a miniscule portion of the discharge energy (~0.0057 %) contributes to an additional ablation driven by the discharge, even if the electrical discharge energies lie in a range between 3.3 to 4.7 J. The cause for such a drastically small energy portion acting during the electrical discharge is likely due to energy dissipation processes during the formation of the electrical discharge arc, the timing when the pulse and the discharge interact with the sample and the relatively long discharge time, which might very likely be on the order of a few microseconds. Since the energy dissipation occurs due to ionization and heating of air, the discharge arrives at the sample at 80 ns or later after the laser pulse, when resolidification processes are already taking place. Even this time (80 ns), the full electrical discharge has not yet been fully built up, only until later than 2 µs. Nevertheless, we observe heating, melting and ablation of the sample caused by the electric discharge.

This begs the question if there might be a way to time the electrical discharge together with a second laser pulse to arrive at effectively the same time so that both interactions may act in concert on the ablation of the sample. Maybe even more important is the question, if there is an electric discharge energy, which allows to utilize a higher portion of the discharge on the sample. This question is inspired by the fact that at some discharge energy nearly all particles in the discharge channel should become ionized. A further increase of the discharge energy would not be able to fuel further ionization. While scattering losses, heat and radiative dissipation would still be present and scale with the discharge energy, the overall efficiency of the discharge on the ablated volume should increase significantly.

So far, our work suggests that the most practicable way to conduct energy-efficient remote ablation of metallic materials seems to use only the laser itself. Note that the laser peak pulse fluence used in this experiment (50 J/cm²) exceeds by far the optimum of the energy specific ablation volume (ESAV) unless the background energy reservoir formed by the process of plasma filamentation is able to redistribute the pulse's energy into a larger area. Therefore, significant increase of the ablated volume (per applied Joule of pulse energy) can be expected if the pulse energy would be reduced. Moreover, as the experiment is being conducted at rather low repetition rates (~23.9 Hz) the material cools down between individual pulses back to room temperature. This stands in contrast to highly efficient ablation processes, where the material is irradiated at such high pulse repetition rates (MHz-rates), that a heat accumulation regime is achieved, which would help to increase the ESAV even further.

## Appendix A – remote principle with filament-guided discharges

Since the aim of this work is to analyze the effect of filament-guided discharges on ablation, we believe it is important to point out that remote processing with such a system is indeed technically feasible. Therefore, we want to present here a schematic that might provide remote random access processing capability in combination with filament-guided discharges.

As a first, it is necessary to note that any laser beam deflection system would need to provide a deflectable plasma filament. The challenge is, however, that the high-voltage electrode has likely too much inertia to be moved fast enough with the plasma filament. Hence, the deflectable plasma filament would need to pass close to (or through) the electrode and then go onto the sample that is to be irradiated. Consequently, a suitable plasma filament deflection system would need to exhibit a pivot point close to the electrode. One potential way to obtain such a pivot point would be to place a fast galvanometric scanner for 2 dimensions into the collimated fs-laser beam, and insert a large focusing lens after the scanner at, e.g., twice its focal length. This would re-image the scanner mirrors after twice the focal length distance and create a pivot point, which might then be placed close to the fixed electrode.

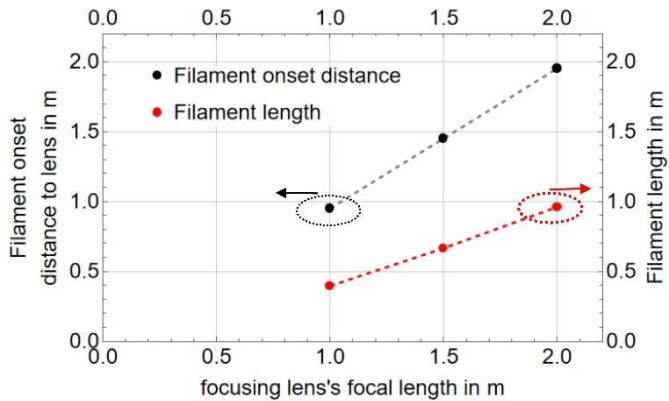

Figure A1: The filament onset distance and filament length measured for different focal lengths. The pulse energy in this measurement was 6.2 mJ, with a collimated beam diameter of 5.5 mm.

However, since the laser generated filament is caused by non-linear self-focusing, see e.g. [2], the filament onset, assuming a collimated beam at the scanner, would occur at a distance less than one focal length of the lens. Unfortunately, in our case the filament length was approximately just half the length of the focal distance of the optics used to generate the filament, compare Fig. A1. Consequently, the filaments would actually end at the pivot point, and not start there, as required by the application.

To overcome this, an optical system similar to the schematic shown in Fig. A2 should be implemented: After passing the scanner, the collimated laser beam is made parallel to the optical axis (just one possible realization – other setups that will have the same capability are possible too) by a lens that is placed one focal length away from the pivot point of the scanner mirrors. Because this lens produces an intermediate focal plane, the corresponding beam segment must propagate through vacuum to avoid pulse-energy decrease and beam deterioration due to nonlinear effects including filamentation. After the beam had expanded sufficiently to avoid filamentation, it exits the vacuum and is refocused by a second lens (lens 2). This converts the now in-parallel propagating beams back to beams that have an intersection with the optical axis, leading effectively to the formation of the desired pivot point.

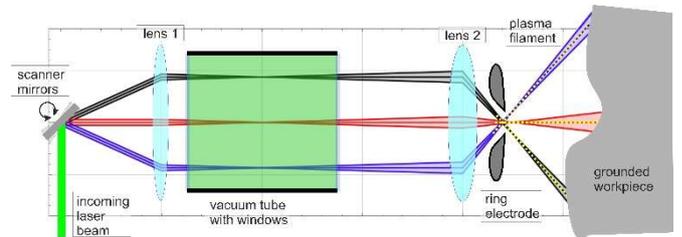

Figure A2: A conceptional optical system allowing for remote processing with laser filament guided electrical discharges with a fixed electrode.

However, care must be taken in choosing the focal length and distance of lens 2 with respect to lens 1. For identical focal lengths of lenses 1 and 2, and a distance of twice the focal length between them, this setup would yield collimated beams, which might lead to a too far onset of the filamentation. Therefore, the distance of lens 2 (to lens 1) must be either larger than twice the focal length or its focal length needs to be shorter – in the ideal case both: longer distance and shorter focal length. In such a case the focal spot "moves" from infinity closer to the pivot point. Nevertheless, care must be taken to not exceed a certain numerical aperture of the beam after lens 2, as this would reduce the filament length. With the focal spot and the pivot point being close together a ring electrode can be placed around the optical axis (compare Appendix B), so that the generated filaments can "connect" the electrode to the grounded work piece.

## Appendix B – effect of electrode shape

Since our previous work shows that the frequency of successful channeling of discharges through the plasma filament is strongly dependent on the direction and position of the electrode [51], a radially symmetrical toroidal geometry of the electrode was thought to be most suitable. Due to the toroidal shape, the electric field should provide a homogeneous distribution for different filament angles with a common pivot point within the electrode, so that all plasma filament directions would exhibit the same probability for channeling. However, a second strong condition was that no undesired discharges occur outside the pivot point, to suppress a voltage drop down of the HV generator during operation. To this end, two different toroid types, see Fig. B1, were designed and manufactured. Here, we present the simulational and experimental results.

The electric field distribution at the electrodes was simulated in order to investigate the ability of the electrodes to channel discharges in a targeted manner using the plasma filament prior to actually manufacturing the electrodes. The electrostatics module from COMSOL Multiphysics was used for this, with the field distribution of the electrodes being simulated in a two-dimensional rotationally symmetrical geometry to save computing time. The grounding electrode was located at a distance of 10 cm from the nearest surface of the toroidal electrode along the axis of rotation and was chosen to be significantly larger than the toroidal electrode itself.

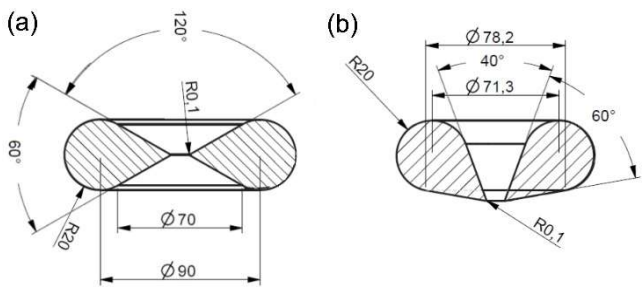

Figure B1: Dimensions of the different electrode types available for the experiment. Type I (a) and type II (b).

The most optimal shape for both electrode types was found to be an outer toroid ring radius of 20 mm that tapers linearly towards the center of the electrode and forms an opening of 10 mm for the laser beam. The edge around the inner opening was also rounded to 0.1 mm so that there are no excessively sharp edges that could lead to unwanted corona discharges.

The main difference between the electrode types was the angle that the linear taper forms with the axis of rotation: while for type I, see Fig. B1 (a), the center axis of the taper is at a right angle to the axis of rotation, the center axis of the electrode type II, Fig. B1 (b), forms an angle of 50° with the axis of rotation. As a result, the pivot point of electrode type I is effectively 20 mm away from the outer edge of the electrode, while for the type II electrode the pivot point coincides directly with the outer edge of the electrode, which, among other things, also maximizes the usable length of the laser filament.

The results for both electrode shapes are shown in Fig. B2. The applied voltage was chosen in both cases (see bottom row in Fig. B2 (c) and (d)) such that the electric field remained below 25 kV/cm at all points on the electrode and its mount, as spontaneous discharges may occur above this threshold [52]. As can be seen, the field strength inside the electrode, i.e. at the pivot point, is significantly lower for type I while being higher on its outer surface torus, whereas for the type II it is exactly the opposite case, as the electric field at the pivot point is significantly higher than on the outer surface of its torus. Since electrode II has a higher field strength near the filament, it can be expected that the plasma filaments can channel discharges easier, i.e. already at lower voltages, and that fewer unchanneled discharges occur on the outer surface of the electrode.

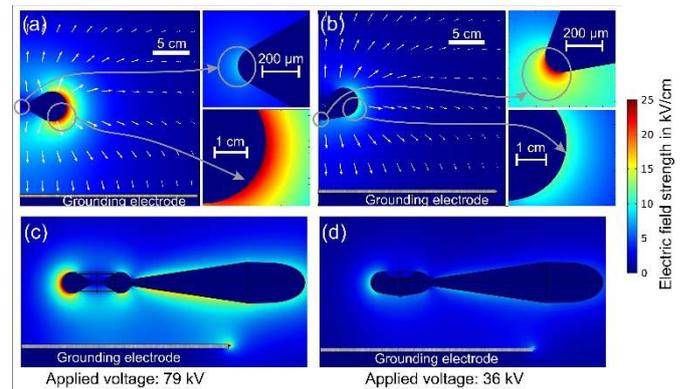

Figure B2: Theoretically calculated electric field strength in the vicinity of the electrode for type I (a) and type II (b). The same calculation is repeated for the electrode including its mount attaching it to the HV generator for the electrode type I (c) and type II (d).

Since the numerical shape optimization, compare Fig. B2 (a) and (b), was carried out in just rotational symmetry, while the electrode needs to be attached to the HV generator, both electrode types were then simulated three-dimensionally together with the holder without the rotational symmetry. The results in Fig. B2 (c) and (d) show an increase of the electric field on the electrode's mount. Nevertheless, the previously calculated field for the type II electrode remains largely as before. While the type II electrode appears to have better properties than type I, we nevertheless tested both electrodes in combination with the USP laser to test their feasibility for laser-filament-guided electrical discharges.

For measuring the free, unguided discharges, the laser was switched off. For each investigated charging voltage we monitored by using cameras and a microphone, whether discharges occurred within the charging time (approx. 30 s). The process was repeated up to 15 times at the same voltage setting. The probability of getting a free discharge at a set voltage was calculated by dividing the number of tests with an observed discharge by the total number of tests run at the voltage. For plasma-filament-guided discharges, the procedure was similar, but with the laser switched on so that filaments were present between the electrodes. In addition, different camera perspectives were used to distinguish between filament-guided discharges and free, unguided discharges. The result is shown in Fig. B3.

As can be seen in Fig. B3, the type II electrode offers more than 50 kV difference between the onset of guided discharges to unguided discharges, while type I offers 16 kV at best. This means, that the break-down voltages are significantly lower in the presence of the plasma filament with the electrode of type II attached, than without the filament or the other electrode. Such a situation is highly desirable, because it allows to conduct ablation experiments with guided discharges while completely avoiding spontaneous free discharges that might occur with the type I electrode. Moreover, because of this, the type II electrode offers a broader range of discharge voltages that can be investigated.

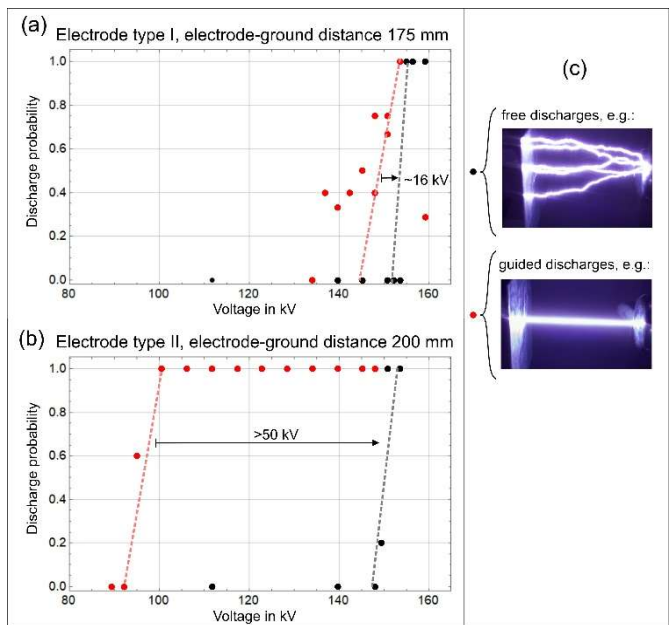

Figure B3: Comparison of the probabilities of a free, unguided discharge and a filament-guided with type I electrode (a) and type II electrode (b). The legend (c) shows in an exemplary way, how the different discharge types were distinguished by means of the cameras (here we show only one perspective).

### Appendix C – crater size

When estimating the crater size using Fig. 5 the ablated craters appear to be roughly the same size for all ablation modes. Taking the dark outline of the most regular crater (c1, c2), i.e. when no discharge and no filament is present, one arrives at a crater radius of ~433 µm or a crater area of 0.59 mm², on the one hand. On the other hand, comparing these size measurements to the visibly ablated depth (bottom row in Fig. 4), shows that the black area must be rather ablation debris. Indeed, careful analysis of the crater height shows actually an elevated ring structure with an increased height of 3.5 µm above the actual sample surface peaking at a radius of 385 µm and fading out at a radius of 500 µm. Therefore, we chose to apply a depth-criterion to estimate the crater area: we determine the crater outline as the rim, where the inner region of the crater falls below the surface level height of the area where debris did not reach, i.e. pristine surface level height. This way, we obtain for the ablation with guided discharges or plasma filament only (i.e. no discharge) area values in the range between 0.4 and 1 mm². For the ablation without discharge and filament, this method provides areas of 0.01 mm² and below. The results are shown in Fig. 5.

### Appendix D – Estimation of energy necessary for the (lossless) ablation of steel 1.3343

Since it was not possible for us to conduct calorimetric measurements during the experiments described in this work, we could not determine the energy directly. Nevertheless, we want to compare the actions of the *filament-only* and the *discharge modes* in terms of efficiency. This comparison was our initial motivation for this project, compare Section 1. Therefore, we need a value for the energy necessary to achieve a phase transition from the solid state at room temperature to the gaseous state of the investigated steel 1.3343.

This steel is an alloy of several elements with vastly differing melt and boiling points, different heat capacities and complex mixing enthalpies. While there exist proprietary programs that implement methods able to deduce the desired material constants, the resulting values might not be as precise as possible. This is because the ablation, melt and evaporation processes occur on time scales between femto- and microseconds, which result in highly non-equilibrium thermodynamic interactions, which are not covered by the programs' methods.

Because of this, we use a very simple model where we calculate the sublimation enthalpy of this steel as the average sublimation enthalpy weighted by portions of the elements. While elements with low boiling point will promote the removal of the melt earlier than calculated by this model, we can see this model as a kind of upper limit for the energy to vaporize the material completely. Consequently, this model also does not include any mixing or phase transition enthalpies.

|  | iron | tungsten | vanadium | molybdenum | chromium | silicon | carbon |
|---|---|---|---|---|---|---|---|
| mass portions in % [15] | 81.575 | 6.3 | 1.9 | 4.95 | 4.15 | 0.225 | 0.9 |
| sublimation enthalpies in kJ/mol [53] | 415.5 | 851 | 515.5 | 658.98 | 397.48 | 450 | 716.67 |
| molar masses in g/mol [54] | 55.845 | 183.84 | 50.942 | 95.95 | 51.996 | 28.085 | 12.011 |
| calculated sublimation enthalpies in kJ/g | 6.069 | 0.292 | 0.192 | 0.340 | 0.317 | 0.036 | 0.537 |

Table D1: Material properties for estimating the sublimation enthalpy of 1.3343 steel.

The mass portions of the elemental constituents of the steel 1.3343, according to [15] are in the first row of Table D1. The corresponding sublimation enthalpies, taken from [53], are in the second row. The respective molar masses, taken from [54], are in the 3rd row. The individual contributions of the elements to the sublimation enthalpy is calculated by dividing the product of the elements' sublimation enthalpies and their mass portions by their respective molar mass. The result is shown in the last row. The sum of the last row gives the total sublimation enthalpy as 7.78 kJ/g. Dividing this value by the steel's density of 8.138 g/cm³ provided e.g. by [55] gives the volumetric sublimation enthalpy of 63.34 kJ/cm³ used in Section 4.3.

## Acknowledgements

The authors gratefully acknowledge support by the DFG, Grant Nos. 263891905 and 447787880, Graduate School in Advanced Optical Technologies (SAOT) of the Friedrich-Alexander-University of Erlangen-Nürnberg, and the Bayerisches Laserzentrum GmbH. Furthermore, the authors want to thank Midhila Krishna for her invaluable help in the experiments and evaluation of the samples.